\documentclass[prd,nofootinbib,preprint, 10 pt]{revtex4}
\usepackage{amssymb}
\usepackage{amsmath,graphicx,color,epsfig}
\usepackage{pstricks}
\usepackage{float}

\begin{document}

\title{Model Independent Analysis of the Forward-Backward Asymmetry for the $%
B\rightarrow K_{1}\mu^{+}\mu^{-}$ Decay}
\author{Ishtiaq Ahmed}
\email{ishtiaq@ncp.edu.pk}
\affiliation{National Centre for Physics, Quaid-i-Azam University, Islamabad, Pakistan}
\author{M. Ali Paracha}
\email{ali@ncp.edu.pk}
\affiliation{Physics Department, Quaid-i-Azam University, Islambad, Pakistan.}
\author{M. Jamil Aslam}
\email{jamil@ncp.edu.pk}
\affiliation{Physics Department, Quaid-i-Azam University, Islamabad, Pakistan}
\date{\today }

\begin{abstract}
The sensitivity of the zero position of the forward backward asymmetry $%
\mathcal{A}_{FB}$ for the exclusive $B\rightarrow K_{1}(1270)\mu^{+}\mu^{-}$
decay is examined by using most general non-standard 4-fermion interactions.
Our analysis shows that the zero position of the forward backward asymmetry
is very sensitive to the sign and size of the Wilson coefficients
corresponding to the new vector type interactions, which are the counter
partners of the usual Standard Model operators but have opposite chirality.
In addition to these, the other significant effect comes from the
interference of Scalar-Psudoscalar and Tensor type operators. These results
will not only enhance our theoretical understanding about the axial vector
mesons but will also serve as a good tool to look for physics beyond the SM.
\end{abstract}

\maketitle


\section{\protect\bigskip Introduction}

Flavor changing neutral current transitions (FCNC) which generally arise at
loop level provides a good testing ground for the Standard Model (SM) \cite%
{1, 2}. Moreover, in such transitions the New Physics (NP) effects can be
probed via the loop of the particles that are beyond the spectrum of SM.
Therefore, there are solid reasons, both theoretical and experimental, for
studying these FCNC transitions. Among all the FCNC processes, the rare $B$
decays are important since one can test both the SM and the possible NP
effects by comparing the theoretical results with the current and future
experiments.

Some of the radiative and semileptonic decays of $B$ mesons to vector and
axial vector mesons, such as $B \to K^{\ast}\gamma$ \cite{1exp,2exp,3exp}, $%
B \to K_{1}(1270, 1400) \gamma$ \cite{4exp} and $B \to
K^{\ast}(892)e^{+}e^{-} (\mu^{+}\mu^{-})$ \cite{5exp, 6exp} have been
observed and for $B \to K^{\ast}(892)e^{+}e^{-} (\mu^{+}\mu^{-})$ the
measurement of isospin and forward-backward asymmetry at BABAR is also
reported \cite{7exp, 8exp, 9exp}. For $B \to K_{1}(1270, 1400) \gamma$ the
Belle has given the following branching frations
\begin{eqnarray}
Br(B \to K_{1}(1270) \gamma)=(4.28 \pm 0.94 \pm 0.43)\times 10^{-5}  \notag
\\
Br(B \to K_{1}(1400) \gamma)<1.44 \times10^{-5}
\end{eqnarray}

The semileptonic $B$ meson decays, $B\rightarrow (K,K^{\ast })l^{+}l^{-}$ ($%
l=e,\mu ,\tau $) are widely studied in the literature \cite{11} where
different physical observables like decay rate, lepton forward-backward
asymmetry and lepton polarizations are calculated both in SM and beyond.
Among these physical observables, the most interesting one is the lepton
forward-backward asymmetry $A_{FB}$ and this lies in the vanishing of $%
A_{FB} $ at a specific value of dilepton mass in a hadronically clean way
\cite{new1,new2,nw1}. This in other words provide a simple relationship
between the electric dipole coefficient $C_{7}$ and $C_{9}$, which is almost
free from the hadronic uncertainties which arises dominantly from the form
factors \cite{new2}.

The above mentioned decays also open a window to look for new Physics. We
know that in SM the decays $B \to (K, K^{*}) l^{+}l^{-}$ are completely
determined by the Wilson coefficients of only three operators $O_{7}$, $%
O_{9} $ and $O_{10}$ which are evaluated at the scale $\mu = m_{b}$ \cite%
{new3}. On the other hand the most general analysis of these decays needs
other set of new operators which are based on the the general four-fermion
interactions. The new structure of effective Hamiltonian \cite{new4, new5}
makes them an ideal platform for the SM, and provide clues for the NP. In
the literature, the model independent analysis of the quark level $b \to s
l^{+} l^{-}$ decay, in terms of 10 new types of local four fermion
interactions, has been performed in Ref. \cite{new4} which is then applied
to the systematic study of $B \to (K, K^{*}) l^{+}l^{-}$ \cite{new6}.
Recently, the discrepancy has been observed in the lepton forward-backward
asymmetry in the exclusive $B\rightarrow K^{\ast }\mu ^{+}\mu ^{-}$ decay
\cite{newref1, newref2}. To explain the experimental results, Kumar et al.
\cite{newref3} have done a systematic study $B\rightarrow K^{\ast }\mu
^{+}\mu ^{-}$ decay by using the most general model independent Hamiltonian.
They have shown that though the scalar and tensor operators are not very
important to study the lepton forward-backward asymmetry but the
interference of these two is important and is not ignorable which differ
from the results given in \cite{new6}.

As the radiative decay $B\rightarrow K_{1}(1270)\gamma $ has already seen by
Belle, therefore the related decay with a lepton pair instead of a photon in
the final state can also be expected to be seen. Analysis of this decay
process will be a useful complement to the widely investigated analysis for
the $B\rightarrow K^{\ast }l^{+}l^{-}$ process, since the analysis probes
the effective Hamiltonian in a similar but not idential way. The
experimental investiagation of this decay will thus provide us independent
test of the predication of the SM and also give us the clue for NP.

Like $B\rightarrow K^{\ast }l^{+}l^{-}$ the semileptonic decay $B\rightarrow
K_{1}(1270)l^{+}l^{-}$ is also governed by the quark level transition $%
b\rightarrow sl^{+}l^{-}$. Compared to $B\rightarrow K^{\ast }l^{+}l^{-}$
the situation is complicated in the decay $B\rightarrow
K_{1}(1270)l^{+}l^{-} $, because the axial vector states $K_{1}(1270)$ and $%
K_{1}(1400)$ are the mixtures of ideal $^{1}P_{1}(K_{1A})$ and $%
^{3}P_{1}(K_{1B})$ orbital angular momentum states and current limit on the
mixing angle is \cite{newr11}
\begin{equation}
\theta =-(34\pm 13)^{o}.
\end{equation}%
Recently, some studies have been made on $B\rightarrow K_{1}$ transitions
both by incorporating the mixing angle as well as with out it \cite{newr12}.

Experimentally, this decay has not yet been seen, but is expected to be
observed at LHC \cite{new8} and SuperB factory \cite{new9}. In particular
LHCb experiment at the LHC where estimates made in \cite{new8, new10} for
LHCb collaboration show that with an integrated luminosity of $2fb^{-1}$,
one may expect almost $8000$ $B\rightarrow K^{\ast }l^{+}l^{-}$ events.
Although the branching ratio of $B\rightarrow K_{1}(1270)l^{+}l^{-}$
calculated in \cite{23} is an order of magnitude smaller than  the
experimentally measured value of $B\rightarrow K^{\ast }l^{+}l^{-}$ \cite%
{PDG}, but still one can expect the significant number of events for this
decay and hence making analysis of FB asymmetry for this decay will be
experimentally meaningful for comparison with the SM and the theories beyond
it.

In this work, our aim is to analyze the possible new physics effects
stemming from the new structures in the effective Hamiltonian \cite{new5} to
the forward-backward asymmetry for the $B\rightarrow K_{1}(1270)l^{+}l^{-}$
decay. It has already been mentioned that some experimental analysis for the
decay $B\rightarrow K^{\ast }\mu ^{+}\mu ^{-}$ has already been studied in $B
$ factories \cite{new9}, but only the large increase in statistics at LHC$b$
for $B\rightarrow K^{\ast }\mu ^{+}\mu ^{-}$ will make much higher precision
measurements possible \cite{new8, new10}. It is known that the
forward-backward asymmetry becomes zero for a particular value of the
dilepton invariant mass. In the SM, the zero of the $\mathcal{A}_{FB}(q^{2})$
appears in the low $q^{2}$ region, sufficiently away from the charm
resonance region and is almost free from the hadronic uncertainties (i.e.
the choice of form factors) and so is from the mixing angle. Now this zero
position of $\mathcal{A}_{FB}$ varies from model to model and this makes it
an important tool to search for physics beyond the SM. The organization of
the paper is as follows: In section II we introduce the model independent
effective Hamiltonian and obtain the transition matrix elements in terms of
form factors of the $B\rightarrow K_{1}(1270)l^{+}l^{-}$. Section III
describes the formulas that can be used to determine the zero position of
the FBA. In Sec. IV we present our numerical analysis and Sec.V summarizes
our conclusion.

\section{Effective Hamiltonian and Matrix Elements}

By integrating out the heavy degrees of freedom in the full theory, the
general effective Hamiltonian for $\ b\rightarrow \ sl^{+}l^{-}$ transitions
in the SM can be written as
\begin{eqnarray}
H_{eff} &=&-\frac{4 G_{F}}{\sqrt{2}}V_{tb}V_{ts}^{\ast }\bigg[{%
\sum\limits_{i=1}^{10}}C_{i}({\mu })O_{i}({\mu })\bigg],
\label{effective haniltonian 1}
\end{eqnarray}
where $O_{i}({\mu })$ $(i=1,\ldots ,10)$ are the four-quark operators and $%
C_{i}({\mu })$ are the corresponding Wilson coefficients at the energy scale
${\mu }$ \cite{new3}. Using renormalization group equations to resum the QCD
corrections, Wilson coefficients are evaluated at the energy scale ${\mu =m}%
_{b}$. The theoretical uncertainties associated with the renormalization
scale can be substantially reduced when the next-to-leading-logarithm
corrections are included.

The explicit expressions of the operators responsible for exclusive $B \to
K_{1}(1270) l^{+}l^{-}$ transition are given by
\begin{eqnarray}
O_{7} &=&\frac{e^{2}}{16\pi ^{2}}m_{b}\left( \bar{s}\sigma _{\mu \nu
}P_{R}b\right) F^{\mu \nu },\  \\
O_{9} &=&\frac{e^{2}}{16\pi ^{2}}(\bar{s}\gamma _{\mu }P_{L}b)(\bar{l}\gamma
^{\mu }l),\  \\
O_{10} &=&\frac{e^{2}}{16\pi ^{2}}(\bar{s}\gamma _{\mu }P_{L}b)(\bar{l}%
\gamma ^{\mu }\gamma _{5}l),  \label{relvent-operators}
\end{eqnarray}
with $P_{L,R}=\left( 1\pm \gamma _{5}\right) /2$. In terms of the above
Hamiltonian, the free quark decay amplitude for $b\rightarrow s$ $l^{+}l^{-}
$ is:

\begin{eqnarray}
\mathcal{M}_{SM}(b &\rightarrow &sl^{+}l^{-})=-\frac{G_{F}\alpha }{\sqrt{2}%
\pi }V_{tb}V_{ts}^{\ast }%
\begin{array}{l}
\bigg\{C_{9}^{eff}(\bar{s}\gamma _{\mu }P_{L}b)(\bar{l}\gamma ^{\mu
}l)+C_{10}(\bar{s}\gamma _{\mu }P_{L}b)(\bar{l}\gamma ^{\mu }\gamma _{5}l)\
-2m_{b}C_{7}^{eff}(\bar{s}i\sigma _{\mu \nu }\frac{q^{\nu }}{s}P_{R}b)(\bar{l%
}\gamma ^{\mu }l)\bigg\}%
\end{array}
\notag  \label{quark-amplitude} \\
&&
\end{eqnarray}%
where $s=q^{2}$ and $q$ is the momentum transfer. The operator $O_{10}$ can
not be induced by the insertion of four-quark operators because of the
absence of the $Z$ boson in the effective theory. Therefore, the Wilson
coefficient $C_{10}$ does not renormalize under QCD corrections and hence it
is independent of the energy scale. In addition to this, the above quark
level decay amplitude can receive contributions from the matrix element of
four-quark operators, $\sum_{i=1}^{6}\langle l^{+}l^{-}s|O_{i}|b\rangle $,
which are usually absorbed into the effective Wilson coefficient $%
C_{9}^{eff}(\mu )$, that one can decompose into the following three parts
\cite{b to s in theory 3,b to s in theory 4,b to s in theory 5,b to s in
theory 6,b to s in theory 7,b to s in theory 8,b to s in theory 9}
\begin{equation*}
C_{9}^{eff}(\mu )=C_{9}(\mu )+Y_{SD}(z,s^{\prime })+Y_{LD}(z,s^{\prime }),
\end{equation*}%
where the parameters $z$ and $s^{\prime }$ are defined as $%
z=m_{c}/m_{b},\,\,\,s^{\prime }=q^{2}/m_{b}^{2}$. $Y_{SD}(z,s^{\prime })$
describes the short-distance contributions from four-quark operators far
away from the $c\bar{c}$ resonance regions, which can be calculated reliably
in the perturbative theory. The long-distance contributions $%
Y_{LD}(z,s^{\prime })$ from four-quark operators near the $c\bar{c}$
resonance cannot be calculated from first principles of QCD and are usually
parameterized in the form of a phenomenological Breit-Wigner formula making
use of the vacuum saturation approximation and quark-hadron duality. The
manifest expressions for $Y_{SD}(z,s^{\prime })$ and $Y_{LD}(z,s^{\prime })$
can be written as
\begin{eqnarray}
Y_{SD}(z,s^{\prime }) &=&h(z,s^{\prime })(3C_{1}(\mu )+C_{2}(\mu
)+3C_{3}(\mu )+C_{4}(\mu )+3C_{5}(\mu )+C_{6}(\mu ))  \notag \\
&&-\frac{1}{2}h(1,s^{\prime })(4C_{3}(\mu )+4C_{4}(\mu )+3C_{5}(\mu
)+C_{6}(\mu ))  \notag \\
&&-\frac{1}{2}h(0,s^{\prime })(C_{3}(\mu )+3C_{4}(\mu ))+{\frac{2}{9}}%
(3C_{3}(\mu )+C_{4}(\mu )+3C_{5}(\mu )+C_{6}(\mu )),
\end{eqnarray}%
with
\begin{eqnarray}
h(z,s^{\prime }) &=&-{\frac{8}{9}}\mathrm{ln}z+{\frac{8}{27}}+{\frac{4}{9}}x-%
{\frac{2}{9}}(2+x)|1-x|^{1/2}\left\{
\begin{array}{l}
\ln \left\vert \frac{\sqrt{1-x}+1}{\sqrt{1-x}-1}\right\vert -i\pi \quad
\mathrm{for}{{\ }x\equiv 4z^{2}/s^{\prime }<1} \\
2\arctan \frac{1}{\sqrt{x-1}}\qquad \mathrm{for}{{\ }x\equiv
4z^{2}/s^{\prime }>1}%
\end{array}%
\right. ,  \notag \\
h(0,s^{\prime }) &=&{\frac{8}{27}}-{\frac{8}{9}}\mathrm{ln}{\frac{m_{b}}{\mu
}}-{\frac{4}{9}}\mathrm{ln}s^{\prime }+{\frac{4}{9}}i\pi \,\,.
\end{eqnarray}%
and
\begin{equation}
Y_{LD}\left( z,s^{\prime }\right) =\frac{3\pi }{\alpha ^{2}}%
C^{(0)}\sum\limits_{V_{i}=\psi _{i}}\kappa _{i}\frac{m_{V_{i}}\Gamma \left(
V_{i}\rightarrow l^{+}l^{-}\right) }{m_{V_{i}}^{2}-s^{\prime
}m_{b}^{2}-im_{V_{i}}\Gamma _{V_{i}}}
\end{equation}%
where $C^{(0)}=3C_{1}+C_{2}+3C_{3}+C_{4}+3C_{5}+C_{6}$. The $Y_{LD}\left(
z,s^{\prime }\right) $ critically depend on the resonance model used to
describe these LD contributions and as such they have uncertainties. But
these uncertainties will hardly effect the zero position of the FB asymmetry
which lies below this charmonium threshold. Keeping in view that there is no
experimental data on $B\to K_{1}(1270)l^{+}l^{-}$,  we have fixed the values
of the phenomenological parameters $\kappa _{i}$ from $B\rightarrow K^{\ast
}l^{+}l^{-}$, which for the resonances $J/\Psi $ and $\Psi ^{^{\prime }}$
are taken to be $\kappa =1.65$ and $\kappa =2.36$, respectively\cite{new2}.

Apart from this, the non-factorizable effects \cite{b to s 1, b to s 2, b to
s 3,NF charm loop} from the charm loop can bring about further corrections
to the radiative $b\rightarrow s\gamma $ transition, which can be absorbed
into the effective Wilson coefficient $C_{7}^{eff}$. Specifically, the
Wilson coefficient $C_{7}^{eff}$ is given by \cite{17}
\begin{equation*}
C_{7}^{eff}(\mu )=C_{7}(\mu )+C_{b\rightarrow s\gamma }(\mu ),
\end{equation*}%
with
\begin{eqnarray}
C_{b\rightarrow s\gamma }(\mu ) &=&i\alpha _{s}\bigg[{\frac{2}{9}}\eta
^{14/23}(G_{1}(x_{t})-0.1687)-0.03C_{2}(\mu )\bigg], \\
G_{1}(x) &=&{\frac{x(x^{2}-5x-2)}{8(x-1)^{3}}}+{\frac{3x^{2}\mathrm{ln}^{2}x%
}{4(x-1)^{4}}},
\end{eqnarray}%
where $\eta =\alpha _{s}(m_{W})/\alpha _{s}(\mu )$, $%
x_{t}=m_{t}^{2}/m_{W}^{2}$, $C_{b\rightarrow s\gamma }$ is the absorptive
part for the $b\rightarrow sc\bar{c}\rightarrow s\gamma $ rescattering and
we have dropped out the tiny contributions proportional to CKM sector $%
V_{ub}V_{us}^{\ast }$. \newline
In addition to the above mentioned currents, the most general form of the
effective Hamiltonian contains 10 local four fermion interactions which can
contribute to the $B\rightarrow K_{1}(1270)l^{+}l^{-}$ decay and these can
be written as:

\begin{eqnarray}
\mathcal{M}_{\text{new}}(b &\rightarrow &sl^{+}l^{-})=\mathcal{M}^{V-A}+%
\mathcal{M}^{S-P}+\mathcal{M}^{T}  \notag \\
\mathcal{M}^{V-A} &=&\frac{G_{F}\alpha }{\sqrt{2}\pi }V_{ts}^{\ast
}V_{tb}\left\{ C_{LL}\overline{s}_{L}\gamma ^{\mu }b_{L}\overline{l}%
_{L}\gamma ^{\mu }l_{L}+C_{LR}\overline{s}_{L}\gamma ^{\mu }b_{L}\overline{l}%
_{R}\gamma ^{\mu }l_{R}+C_{RL}\overline{s}_{R}\gamma ^{\mu }b_{R}\overline{l}%
_{L}\gamma ^{\mu }l_{L}+C_{RR}\overline{s}_{R}\gamma ^{\mu }b_{R}\overline{l}%
_{R}\gamma ^{\mu }l_{R}\right\}  \notag \\
\mathcal{M}^{S-P} &=&\frac{G_{F}\alpha }{\sqrt{2}\pi }V_{ts}^{\ast
}V_{tb}C_{LRLR}\overline{s}_{L}b_{R}\overline{l}_{L}l_{R}+C_{RLLR}\overline{s%
}_{R}b_{L}\overline{l}_{L}l_{R}+C_{LRRL}\overline{s}_{L}b_{R}\overline{l}%
_{R}l_{L}+C_{RLRL}\overline{s}_{R}b_{L}\overline{l}_{R}l_{L}  \notag \\
\mathcal{M}^{T} &=&\frac{G_{F}\alpha }{\sqrt{2}\pi }V_{ts}^{\ast
}V_{tb}\left\{ C_{T}\overline{s}\sigma _{\mu \nu }b\overline{l}\sigma ^{\mu
\nu }l+iC_{TE}\epsilon _{\mu \nu \alpha \beta }\overline{l}\sigma ^{\mu \nu
}l\overline{s}\sigma ^{\alpha \beta }b\right\}  \label{nphysoperator}
\end{eqnarray}%
Thus the explicit form of the free quark amplitude $\mathcal{M}$ for the $%
b\rightarrow sl^{+}l^{-}$ transition can be written as sum of the SM
amplitude (Eq. (\ref{quark-amplitude})) and of the new physics contributions
(Eq. (\ref{nphysoperator})), i.e.
\begin{equation}
\mathcal{M}=\mathcal{M}_{SM}+\mathcal{M}_{new}  \label{combined}
\end{equation}%
\newline
The exclusive $B\rightarrow K_{1}(1270)l^{+}l^{-}$ decay involves the
hadronic matrix elements of quark operators given in Eq. (\ref%
{quark-amplitude}) and Eq. (\ref{nphysoperator}) which one can be
parametrize in terms of the form factors as follows:
\begin{eqnarray}
\left\langle K_{1}(k,\varepsilon )\left\vert V_{\mu }\right\vert
B(p)\right\rangle &=&\varepsilon _{\mu }^{\ast }\left(
M_{B}+M_{K_{1}}\right) V_{1}(s)  \notag \\
&&-(p+k)_{\mu }\left( \varepsilon ^{\ast }\cdot q\right) \frac{V_{2}(s)}{%
M_{B}+M_{K_{1}}}  \notag \\
&&-q_{\mu }\left( \varepsilon \cdot q\right) \frac{2M_{K_{1}}}{s}\left[
V_{3}(s)-V_{0}(s)\right]  \label{6} \\
\left\langle K_{1}(k,\varepsilon )\left\vert A_{\mu }\right\vert
B(p)\right\rangle &=&\frac{2i\epsilon _{\mu \nu \alpha \beta }}{%
M_{B}+M_{K_{1}}}\varepsilon ^{\ast \nu }p^{\alpha }k^{\beta }A(s)  \label{7}
\end{eqnarray}%
where $V_{\mu }=\bar{s}\gamma _{\mu }b$ and $A_{\mu }=\bar{s}\gamma _{\mu
}\gamma _{5}b$ are the vectors and axial vector currents respectively. Also $%
p(k)$ are the momentum of the $B(K_{1})$ meson and $\varepsilon _{\mu
}^{\ast }$ is the polarization of the final state axial vector $K_{1}$
meson. In Eq.(\ref{6}) we have
\begin{equation}
V_{3}(s)=\frac{M_{B}+M_{K_{1}}}{2M_{K_{1}}}V_{1}(s)-\frac{M_{B}-M_{K_{1}}}{%
2M_{K_{1}}}V_{2}(s)  \label{8}
\end{equation}%
with
\begin{equation*}
V_{3}(0)=V_{0}(0)
\end{equation*}%
In addition to the above, there is also a contribution from the Penguin form
factors that can be written as
\begin{eqnarray}
\left\langle K_{1}(k,\varepsilon )\left\vert \bar{s}i\sigma _{\mu \nu
}q^{\nu }b\right\vert B(p)\right\rangle &=&\left[ \left(
M_{B}^{2}-M_{K_{1}}^{2}\right) \varepsilon _{\mu }-(\varepsilon \cdot
q)(p+k)_{\mu }\right] F_{2}(s)  \notag \\
&&+(\varepsilon ^{\ast }\cdot q)\left[ q_{\mu }-\frac{s}{%
M_{B}^{2}-M_{K_{1}}^{2}}(p+k)_{\mu }\right] F_{3}(s)  \label{9} \\
&&  \notag \\
\left\langle K_{1}(k,\varepsilon )\left\vert \bar{s}i\sigma _{\mu \nu
}q^{\nu }\gamma _{5}b\right\vert B(p)\right\rangle &=&-i\epsilon _{\mu \nu
\alpha \beta }\varepsilon ^{\ast \nu }p^{\alpha }k^{\beta }F_{1}(s)
\label{10}
\end{eqnarray}%
with $F_{1}(0)=2F_{2}(0).$

By contracting Eq. (\ref{6}) with $q_{\mu }$ and making use of the equation
of motions
\begin{eqnarray}
q^{\mu }(\bar{\psi}_{1}\gamma _{\mu }\psi _{2}) =(m_{2}-m_{1})\bar{\psi}%
_{1}\psi _{2}  \label{eq-motion1} \\
q^{\mu }(\bar{\psi}_{1}\gamma _{\mu }\gamma _{5}\psi _{2}) =-(m_{1}+m_{2})%
\bar{\psi}_{1}\gamma _{5 }\psi _{2}  \label{eq-motion}
\end{eqnarray}
we have
\begin{eqnarray}
\langle K_{1}(k,\varepsilon)|\bar{s}(1 \pm \gamma _{5})b|B(p)\rangle = \frac{%
1}{m_{b}+m_{s}}\left\{ \mp 2iM_{K_{1}}(\varepsilon ^{*}\cdot
q)V_{0}(s)\right\}  \label{12}
\end{eqnarray}

The form factors for $B\rightarrow K_{1}(1270)$ transition are the
non-perturbative quantities and are needed to be calculated using different
approaches (both perturbative and non-perturbative) like Lattice QCD, QCD
sum rules, Light Cone sum rules, etc. As the zero position of the
forward-backward asymmetry depends on the short distance contribution i.e.
the Wilson coefficients and is not very sensitive to the long distance
contribution (Form factors) \cite{23} and consequently on the mixing angle
between $^{1}P_{1}$ and $^{3}P_{1}$ states. As such we will consider the
form factors that were calculated using Ward Identities in Ref. \cite{23}
which can be summarized as follows:
\begin{eqnarray}
A(\hat{s}) &=&\frac{A(0)}{\left( 1-\hat{s}\right) }(1-\hat{s}\frac{M_{B}^{2}%
}{M_{B}^{\prime 2}})  \notag \\
V_{1}(\hat{s}) &=&\frac{V_{1}(0)}{(1-\hat{s}\frac{M_{B}^{2}}{M_{B_{A}^{\ast
}}^{2}})(1-\hat{s}\frac{M_{B}^{2}}{M_{B_{A}^{\ast }}^{\prime 2}})}\left( 1-%
\frac{\hat{s}}{1-\hat{M}_{K_{1}}^{2}}\right)   \notag \\
V_{2}(\hat{s}) &=&\frac{\tilde{V}_{2}(0)}{(1-\hat{s}\frac{M_{B}^{2}}{%
M_{B_{A}^{\ast }}^{2}})(1-\hat{s}\frac{M_{B}^{2}}{M_{B_{A}^{\ast }}^{\prime
2}})}-\frac{2\hat{M}_{K_{1}}}{1-\hat{M}_{K_{1}}}\frac{V_{0}(0)}{(1-\hat{s}%
)(1-\hat{s}\frac{M_{B}^{2}}{M_{B}^{\prime 2}})}  \notag \\
&&  \label{13}
\end{eqnarray}%
with
\begin{eqnarray}
V_{0}(0) &=&0.36\pm 0.03 \\
A(0) &=&-(0.52\pm 0.05)  \notag \\
V_{1}(0) &=&-(0.24\pm 0.02)  \notag \\
\tilde{V}_{2}(0) &=&-(0.39\pm 0.05)  \notag \\
&&  \label{14}
\end{eqnarray}

\section{Forward backward asymmetry for $B \rightarrow K_1(1270)l^{+}l^{-}$}

In this section, we are going to perform the calculation of the
forward-backward asymmetry. From Eq. (\ref{quark-amplitude}), it is
straightforward to obtain the decay amplitude for $B\rightarrow
K_{1}(1270)l^{+}l^{-}$ as
\begin{equation}
\mathcal{M}_{B\rightarrow K_{1}(1270)l^{+}l^{-}}=\frac{G_{F}\alpha }{4\sqrt{2%
}\pi }V_{tb}V_{ts}^{\ast }M_{B}\left\{
\begin{array}{c}
T_{\mu }^{1}\overline{l}\gamma ^{\mu }l+T_{\mu }^{2}\overline{l}\gamma ^{\mu
}\gamma ^{5}l+T^{3}\overline{l}l+T^{4}\overline{l}\gamma ^{5}l \\
+8C_{T}(\overline{l}\sigma ^{\mu \nu }l)(-2F_{1}\left( \hat{s}\right)
\varepsilon ^{\ast \mu }\left( \hat{p}_{B}+\hat{p}_{K_{1}}\right) ^{\mu
}+J_{1}\varepsilon ^{\ast \mu }\widehat{q}^{\nu }-J_{2}\left( \varepsilon
^{\ast }\cdot \widehat{q}\right) \hat{p}_{K_{1}}^{\mu }\widehat{q}^{\nu })
\\
+2iC_{TE}\epsilon _{\mu \nu \alpha \beta }(\overline{l}\sigma ^{\mu \nu
}l)(-2F_{1}\left( \hat{s}\right) \varepsilon ^{\ast \alpha }\left( \hat{p}%
_{B}+\hat{p}_{K_{1}}\right) ^{\beta }+J_{1}\varepsilon ^{\ast \alpha }%
\widehat{q}^{\beta }-J_{2}\left( \varepsilon ^{\ast }\cdot \widehat{q}%
\right) \hat{p}_{K_{1}}^{\alpha }\widehat{q}^{\beta })%
\end{array}%
\right\}  \label{15}
\end{equation}%
where the functions $T_{\mu }^{1}$, $T_{\mu }^{2}$, $T^{3}$ and $T^{4}$ in
terms of auxiliary functions are given by
\begin{eqnarray}
T_{\mu }^{1} &=&iA^{\prime }(\hat{s})\epsilon _{\mu \rho \alpha \beta
}\epsilon ^{\ast \rho }\hat{p}_{B}^{\alpha }\hat{p}_{K_{1}}^{\beta
}-B^{\prime }(\hat{s})\epsilon _{\mu }^{\ast }+C^{\prime }(\hat{s}%
)(\varepsilon ^{\ast }\cdot \hat{p}_{B})\hat{p}_{h\mu }+D^{\prime }(\hat{s}%
)(\varepsilon ^{\ast }\cdot \hat{p}_{B})\hat{q}_{\mu }  \notag \\
T_{\mu }^{2} &=&iE^{\prime }(\hat{s})\epsilon _{\mu \rho \alpha \beta
}\epsilon ^{\ast \rho }\hat{p}_{B}^{\alpha }\hat{p}_{K_{1}}^{\beta
}-F^{\prime }(\hat{s})\epsilon _{\mu }^{\ast }+G^{\prime }(\hat{s}%
)(\varepsilon ^{\ast }\cdot \hat{p}_{B})\hat{p}_{h\mu }+H^{\prime }(\widehat{%
s})(\varepsilon ^{\ast }\cdot \hat{p}_{B})\widehat{q}_{\mu }  \notag \\
T^{3} &=&iI^{^{\prime }}(\varepsilon ^{\ast }\cdot \widehat{q})  \notag \\
T^{4} &=&iJ^{^{\prime }}(\varepsilon ^{\ast }\cdot \widehat{q})  \label{16}
\end{eqnarray}
where $\hat{s}=s/M_{B}^{2},$ $\hat{p}_{K_{1}}=p_{K_{1}}/M_{B},$ $\hat{p}%
_{B}=p_{B}/M_{B}$, $\hat{m}_{b}=m_{b}/M_{B}$ and $\hat{M}_{K_{1}}=M_{K_{1}}/M_{B}$.

Defining the combinations
\begin{eqnarray}
C_{RR}^{(+)} &=&C_{RR}+C_{RL},\text{ }C_{RR}^{(-)}=C_{RR}-C_{RL},  \notag \\
C_{LL}^{(+)} &=&C_{LL}+C_{LR},\text{ }C_{LL}^{(-)}=C_{LL}-C_{LR},
\label{combination-coefficients} \\
C_{RLLR}^{(+)} &=&C_{RLLR}+C_{RLRL},~C_{LRRL}^{(+)}=C_{LRRL}+C_{LRLR},
\notag \\
C_{RLLR}^{(-)} &=&C_{RLLR}-C_{RLRL},~C_{LRRL}^{(-)}=C_{LRLR}-C_{LRRL},
\notag
\end{eqnarray}

the auxiliary functions appearing in Eq. (\ref{16}) can be written as
follows:
\begin{eqnarray}
A^{\prime }(\hat{s}) &=&-\frac{2}{1+\hat{M}_{K_{1}}}[C_{9}^{eff}+\frac{1}{2}%
(C_{RR}^{(+)}+C_{LL}^{(+)})]A(\hat{s})+\frac{2\hat{m}_{b}}{\hat{s}}%
C_{7}^{eff}F_{1}(\hat{s})  \notag \\
B^{\prime }(\hat{s}) &=&(1+\hat{M}_{K_{1}})(C_{9}^{eff}+\frac{1}{2}%
(C_{LL}^{(+)}-C_{RR}^{(-)}))V_{1}(\widehat{s})+\frac{2\hat{m}_{b}}{\hat{s}}%
(1-\hat{M}_{K_{1}}^{2})C_{7}^{eff}F_{2}(\widehat{s})  \notag \\
C^{\prime }(\hat{s}) &=&\frac{1}{(1-\hat{M}_{K_{1}}^{2})}\left[ ((1-\hat{M}%
_{K_{1}})(C_{9}^{eff}+\frac{1}{2}(C_{LL}^{(+)}-C_{RR}^{(+)})))V_{2}(\hat{s}%
)+2\hat{m}_{b}C_{7}^{eff}(F_{3}(\hat{s})-(1-\hat{M}_{K_{1}}^{2})/\hat{s}%
)F_{2}(\hat{s})\right]  \notag \\
D^{\prime }(\hat{s}) &=&\frac{1}{\widehat{s}}\left[ ((1+\hat{M}%
_{K_{1}})V_{1}(\hat{s})-(1-\hat{M}_{K_{1}})V_{2}(\hat{s})-2\hat{M}%
_{K_{1}}V_{0}(\hat{s}))(C_{9}^{eff}+\frac{1}{2}(C_{LL}^{(+)}-C_{RR}^{(+)}))-2%
\hat{m}_{b}C_{7}^{eff}F_{3}(\hat{s})\right]  \notag \\
E^{\prime }(\hat{s}) &=&\frac{-2}{1+\hat{M}_{K_{1}}}[C_{10}+\frac{1}{2}%
(C_{RR}^{(-)}-C_{LL}^{(-)})]A(\hat{s})  \notag \\
F^{\prime }(\hat{s}) &=&(1+\hat{M}_{K_{1}})[C_{10}-\frac{1}{2}%
(C_{LL}^{(-)}+C_{RR}^{(-)}]V_{1}(\hat{s})  \notag \\
G^{\prime }(\hat{s}) &=&-\frac{1}{\left( 1+\hat{M}_{K_{1}}\right) }[C_{10}-%
\frac{1}{2}(C_{LL}^{(-)}+C_{RR}^{(-)}]V_{2}(\hat{s})  \notag \\
H^{\prime }(\hat{s}) &=&\frac{1}{\hat{s}}\left[ ((1-\hat{M}_{K_{1}})V_{2}(%
\widehat{s})-(1+\hat{M}_{K_{1}})V_{1}(\widehat{s})+2\hat{M}_{K_{1}}V_{0}(%
\widehat{s}))(C_{10}-\frac{1}{2}(C_{RR}^{(-)}+C_{LL}^{(-)})\right]  \notag \\
I^{^{\prime }}(\hat{s}) &=&\frac{2\hat{M}_{K_{1}}}{\widehat{m}_{b}}V_{0}(%
\widehat{s})[C_{RLLR}^{(+)}+C_{LRRL}^{(+)}]  \notag \\
J^{^{\prime }}(\hat{s}) &=&\frac{2\hat{M}_{K_{1}}}{\hat{m}_{b}}V_{0}(\hat{s}%
)[C_{RLLR}^{(+)}-C_{LRRL}^{(+)}]  \notag \\
J_{1}^{\prime }(\hat{s}) &=&2\left( 1-\hat{M}_{K_{1}}^{2}\right) \frac{%
F_{1}\left( \hat{s}\right) -F_{2}\left( \hat{s}\right) }{\hat{s}}  \notag \\
J_{2}^{\prime }(\hat{s}) &=&\frac{4M_{B}^{2}}{\hat{s}}\left( F_{1}\left(
\hat{s}\right) -F_{2}\left( \hat{s}\right) -\frac{\hat{s}}{1-\hat{M}%
_{K_{1}}^{2}}F_{3}\left( \hat{s}\right) \right)  \label{auxiliary-functions}
\end{eqnarray}%
where, $A^{\prime },~B^{\prime },$ $C^{\prime },~D^{\prime }$, $E^{\prime }\,
$, $F^{\prime }$, $G^{\prime }$, $H^{\prime }$ corresponds to $VA$
interactions where as $I^{\prime }$, $J^{\prime }$ $,$ $J_{1}^{\prime }$, $%
J_{2}^{\prime }$ are relevant for $SP$ and $T$ interactions.

To calculate the forward-backward asymmetry of the final state leptons, one
needs to know the differential decay width of $B\rightarrow
K_{1}(1270)l^{+}l^{-}$, which in the rest frame of $B$ meson can be written
as
\begin{equation}
{\frac{d\Gamma (B\rightarrow K_{1}(1270)l^{+}l^{-})}{ds}}={\frac{1}{(2\pi
)^{3}}}{\frac{1}{32M_{B}}}\int_{u_{min}}^{u_{max}}|\mathcal{M}_{B\rightarrow
K_{1}(1270)l^{+}l^{-}}|^{2}du,  \label{differential decay width}
\end{equation}%
where $u=(k+p_{l^{-}})^{2}$ and $s=(p_{l^{+}}+p_{l^{-}})^{2}$; $k$, $%
p_{l^{+}}$ and $p_{l^{-}}$ are the four-momenta vectors of $K_{1}(1270)$, $%
l^{+}$ and $l^{-}$ respectively; $|\mathcal{M}_{B\rightarrow
K_{1}(1270)l^{+}l^{-}}|^{2}$ is the squared decay amplitude after
integrating over the angle between the lepton $l^{-}$ and $K_{1}(1270)$
meson. The upper and lower limits of $u$ are given by
\begin{eqnarray}
u_{max} &=&(E_{K_{1}(1270)}^{\ast }+E_{l^{-}}^{\ast })^{2}-(\sqrt{%
E_{K_{1}(1270)}^{\ast 2}-M_{K_{1}(1270)}^{2}}-\sqrt{E_{l^{-}}^{\ast
2}-m_{l^{-}}^{2}})^{2},  \notag \\
u_{min} &=&(E_{K_{1}(1270)}^{\ast }+E_{l^{-}}^{\ast })^{2}-(\sqrt{%
E_{K_{1}(1270)}^{\ast 2}-M_{K_{1}(1270)}^{2}}+\sqrt{E_{l^{-}}^{\ast
2}-m_{l^{-}}^{2}})^{2};
\end{eqnarray}%
where $E_{K_{1}(1270)}^{\ast }$ and $E_{l^{-}}^{\ast }$ are the energies of $%
K_{1}(1270)$ and $l^{-}$ in the rest frame of lepton pair and can be
determined as
\begin{equation}
E_{K_{1}(1270)}^{\ast }={\frac{M_{B}^{2}-M_{K_{1}(1270)}^{2}-s}{2\sqrt{s}}},%
\hspace{1cm}E_{l}^{\ast }={\frac{s}{2\sqrt{s}}}.
\end{equation}%
The differential FBA of final state lepton for the said decay can be written
as
\begin{equation}
{\frac{dA_{FB}(s)}{ds}}=\int_{0}^{1}d\cos \theta {\frac{d^{2}\Gamma (s,\cos
\theta )}{dsd\cos \theta }}-\int_{-1}^{0}d\cos \theta {\frac{d^{2}\Gamma
(s,\cos \theta )}{dsd\cos \theta }}
\end{equation}%
and
\begin{equation}
A_{FB}(s)={\frac{\int_{0}^{1}d\cos \theta {\frac{d^{2}\Gamma (s,\cos \theta )%
}{dsd\cos \theta }}-\int_{-1}^{0}d\cos \theta {\frac{d^{2}\Gamma (s,\cos
\theta )}{dsd\cos \theta }}}{\int_{0}^{1}d\cos \theta {\frac{d^{2}\Gamma
(s,\cos \theta )}{dsd\cos \theta }}+\int_{-1}^{0}d\cos \theta {\frac{%
d^{2}\Gamma (s,\cos \theta )}{dsd\cos \theta }}}}.
\end{equation}%
Now putting everything together in hat notation we have
\begin{equation}
\frac{d\mathit{A}_{FB}}{d\hat{s}}=\frac{G_{F}^{2}\alpha ^{2}m_{B}^{5}}{%
2^{10}\pi ^{5}}\left\vert V_{ts}^{\ast }V_{tb}\right\vert ^{2}u(\hat{s})%
\left[ X_{VA}+X_{SP}+X_{T}+X_{VA-SP}+X_{VA-T}+X_{SP-T}\right]
\end{equation}%
where
\begin{eqnarray*}
u(\hat{s}) &=&\sqrt{\lambda (1,\hat{M_{K_{1}}},\hat{s})(1-4\frac{\hat{m}%
_{l}^{2}}{\hat{s}})} \\
\lambda (1,\hat{M}_{K_{1}}^{2},\hat{s}) &=&1+\hat{M}_{K_{1}}^{4}+\hat{s}%
^{2}-2\hat{s}-2\hat{M}_{K_{1}}^{2}(1+\hat{s})
\end{eqnarray*}%
and%
\begin{eqnarray}
X_{VA} &=&M_{B}\hat{s}\hat{M}_{K_{1}}\Re \lbrack A^{\prime \ast }F^{\prime
}+B^{\prime \ast }E^{\prime }]  \notag \\
X_{SP} &=&0  \notag \\
X_{T} &=&0  \notag \\
X_{SP-VA} &=&\hat{m}_{l}\left[ \left( \hat{M}_{K_{1}}^{2}+\hat{s}-1\right)
\Re (B^{\prime \ast }I^{\prime })+M_{B}^{2}\lambda \Re (I^{\prime \ast
}C^{\prime })\right]  \notag \\
X_{SP-T} &=&M_{B}\hat{M}_{K_{1}}^{2}\Re \lbrack 2I^{\prime \ast
}C_{T}+J^{\prime \ast }C_{TE})\left( 2J_{1}^{\prime }(\hat{M}_{K_{1}}^{2}+%
\hat{s}-1)+J_{2}^{\prime }M_{B}^{2}\lambda +4F_{1}\left( \hat{s}\right) (3%
\hat{M}_{K_{1}}^{2}-\hat{s}+1)\right)  \label{expressions-FBA} \\
X_{VA-T} &=&\hat{m}_{l}[2\Re (F^{\prime \ast }C_{TE})\left( 2J_{1}^{\prime }(%
\hat{M}_{K_{1}}^{2}+\hat{s}-1)+J_{2}^{\prime }M_{B}^{2}\lambda +F_{1}\left(
\hat{s}\right) (4\hat{M}_{K_{1}}^{2}-4\hat{s}+4)\right)  \notag \\
&&-2\Re (G^{\prime \ast }C_{TE})M_{B}^{2}\left(
\begin{array}{c}
2J_{1}^{\prime }(\hat{M}_{K_{1}}^{2}\hat{s}-\hat{s}^{2}+\hat{s}+\lambda
)+J_{2}^{\prime }M_{B}^{2}(\hat{M}_{K_{1}}^{2}-1)\lambda \\
+4F_{1}\left( \hat{s}\right) (5\hat{M}_{K_{1}}^{2}\hat{s}+4\hat{M}%
_{K_{1}}^{2}-3\hat{s}^{2}+7\hat{s}+3\lambda -4)%
\end{array}%
\right)  \notag \\
&&+2\Re (H^{\prime \ast }C_{TE})M_{B}^{2}\hat{M}_{K_{1}}^{2}\left(
2J_{1}^{\prime }(\hat{M}_{K_{1}}^{2}+\hat{s}-1)+J_{2}^{\prime
}M_{B}^{2}\lambda +4F_{1}\left( \hat{s}\right) (3\hat{M}_{K_{1}}^{2}-\hat{s}%
+1)\right)  \notag \\
&&-64\Re (E^{\prime \ast }C_{T})M_{B}^{2}\left( J_{1}\hat{M}_{K_{1}}^{2}\hat{%
s}+2F_{1}\left( \hat{s}\right) \left( \hat{M}_{K_{1}}^{2}\hat{s}+\hat{s}-(%
\hat{s}-1)^{2}+\lambda \right) \right) ]  \notag
\end{eqnarray}%
From experimental point of view the normalized forward-backward asymmetry is
more useful, i.e.
\begin{equation*}
\frac{d\mathcal{\bar{A}}_{FB}}{d\hat{s}}=\frac{d\mathcal{A}_{FB}}{d\hat{s}}/%
\frac{d\Gamma }{d\hat{s}}
\end{equation*}

\section{Numerical Analysis}

In the following section, we examine the lepton forward-backward asymmetry
and study the sensitivity of its zero position to New Physics operators. We
consider different Lorentz structures of NP, as well as their combinations
and take all the NP couplings to be real.

\textbf{Switching off all New Physics Operators}

By switching off all the new physics operators one will get the SM result of
the lepton forward-backward asymmetry for $B\rightarrow K_{1}(1270)\mu
^{+}\mu ^{-} $ which was earlier calculated by Paracha et al. \cite{23} and
has been shown by solid line in all the figures shown below. The zero
position lies at $\hat{s}=0.16$ ($s=4.46$ GeV$^{-2}$) and is almost
independent of the choice of form factors and also from the uncertainties
arising from different input parameters like form factors, CKM matrix
elements, etc. In the subsequent analysis we will ignore these uncertanities.

In case of $B\rightarrow K^{\ast }$, Arda et. al. have shown \cite{16} that
the presence of the tensor and the scalar type interactions have very mild
effect on the zero position of forward-backward asymmetry ($\mathcal{A}_{FB}$%
) and they have ignored it in their analysis. However, recently the
discrepancy has been observed in the lepton forward-backward asymmetry in
the exclusive $B\rightarrow K^{\ast }\mu ^{+}\mu ^{-}$ decay \cite{newref1,
newref2}. To explain the experimental results, Kumar et al. \cite{newref3}
have done a systematic study of $B\rightarrow K^{\ast }\mu ^{+}\mu ^{-}$
decay by using the most general model independent Hamiltonian. They have
shown that though the scalar and tensor operators are not important to study
the lepton forward-backward asymmetry but the interference of these two is
important and is not ignorable. Therefore, keeping this in view we will not
ignore these scalar and tensor type couplings in our analysis of $%
B\rightarrow K_{1}(1270)$ decay. In order to see the effect of the new
vector type Wilson coefficients ($%
C_{X}=C_{LL},C_{LR},C_{RR},C_{RL},C_{LRLR},~C_{T},~C_{TE}),$ we have plotted
the dependence of $\mathcal{A}_{FB}$ on $\hat{s}$ by using different values
of $C_{X}$, which can be summarized as follows.

\textbf{Switching on only }$C_{LL}$\textbf{\ and }$C_{LR}$\textbf{\ along
with\ SM\ operators}

Considering the constraints provided by Kumar et al. \cite{newref3} we took
broad range of the values of different VA couplings. Fig. 1(a, b) shows the
dependence of $\mathcal{A}_{FB}$ on $\hat{s}$ when all the $\mathbf{C}_{LL}$
and $C_{LR}$ are present. When $C_{LL(LR)}=-C_{10}$, $C_{LL\left( LR\right)
}=C_{10},$ $C_{LL(LR)}=-0.7\times C_{10},$ $C_{LL\left( LR\right)
}=0.7\times C_{10}$ (and all other Wilson coefficients are set to zero) we
denote the curves of $\mathcal{A}_{FB}$ by dashed double dotted, dashed
triple dotted, dashed and dashed dotted lines respectively. The solid line
corresponds to the SM result. One can deduce from here that there is a
significant shift in the zero position of the forward-backward asymmetry and
the position of zero is gradually shifted to the left for positive values of
$C_{10}$ and to the right for negative values of $C_{10}$ compared to the SM
value. This is contrary to the $B\to K^{*} \mu^{+}\mu^{-}$ decay process
where for the positive values of $C_{LL}(LR)$ the zero position of $\mathcal{%
A}_{FB}$ shifts to the right and for negative value of these new
coefficients the shift in the zero position is to the left \cite{naveen05}.
This difference is due to the axial vector nature of the $K_1(1270)$. For
different values of NP coefficients, the location of the zero of the $%
\mathcal{A}_{FB}$ varies from $\hat{s}$ = $\ 0.12$ to $0.23.$

\begin{figure}[h]
\begin{center}
\begin{tabular}{ccc}
\vspace{-2cm} \includegraphics[scale=0.6]{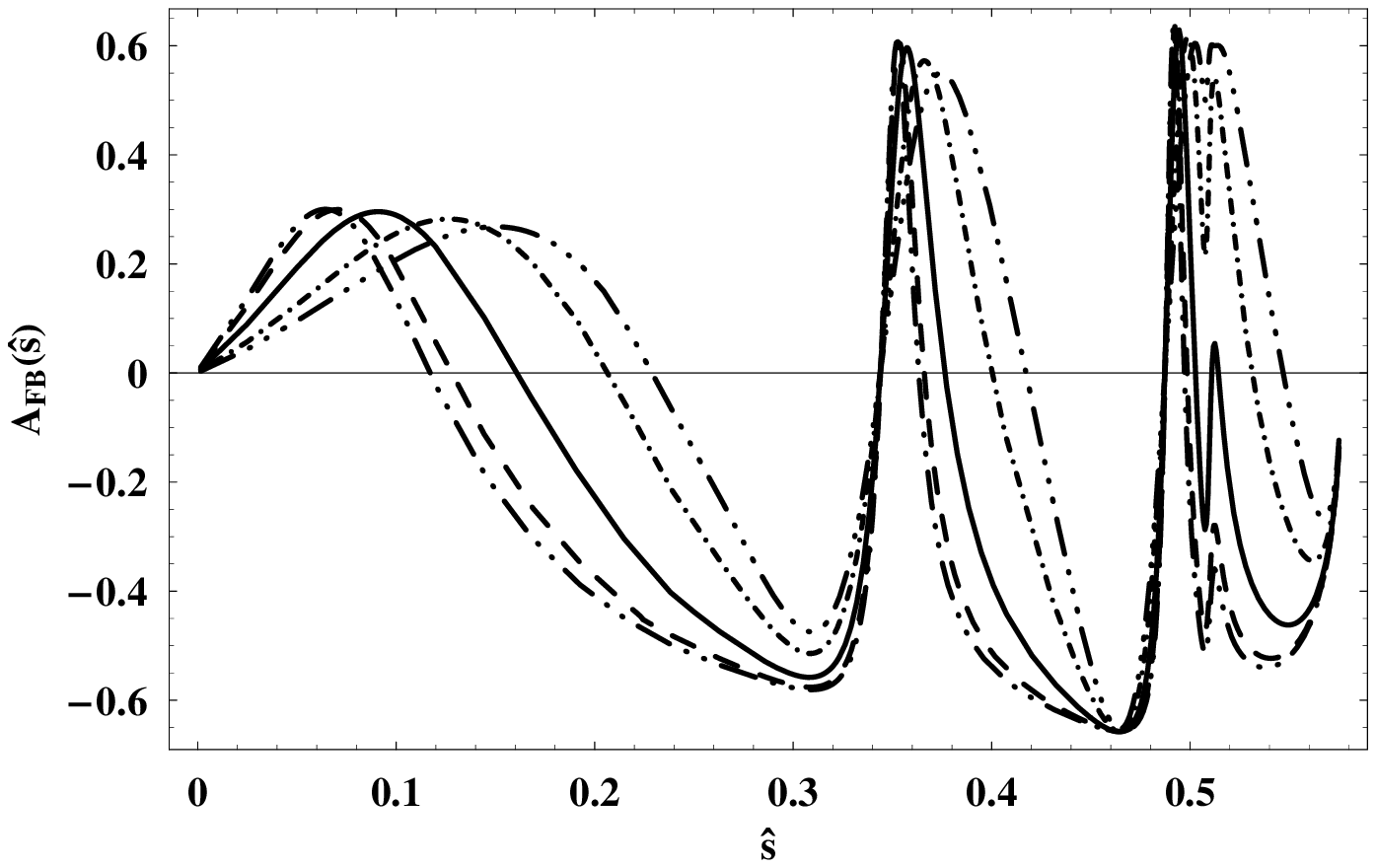} %
\includegraphics[scale=0.6]{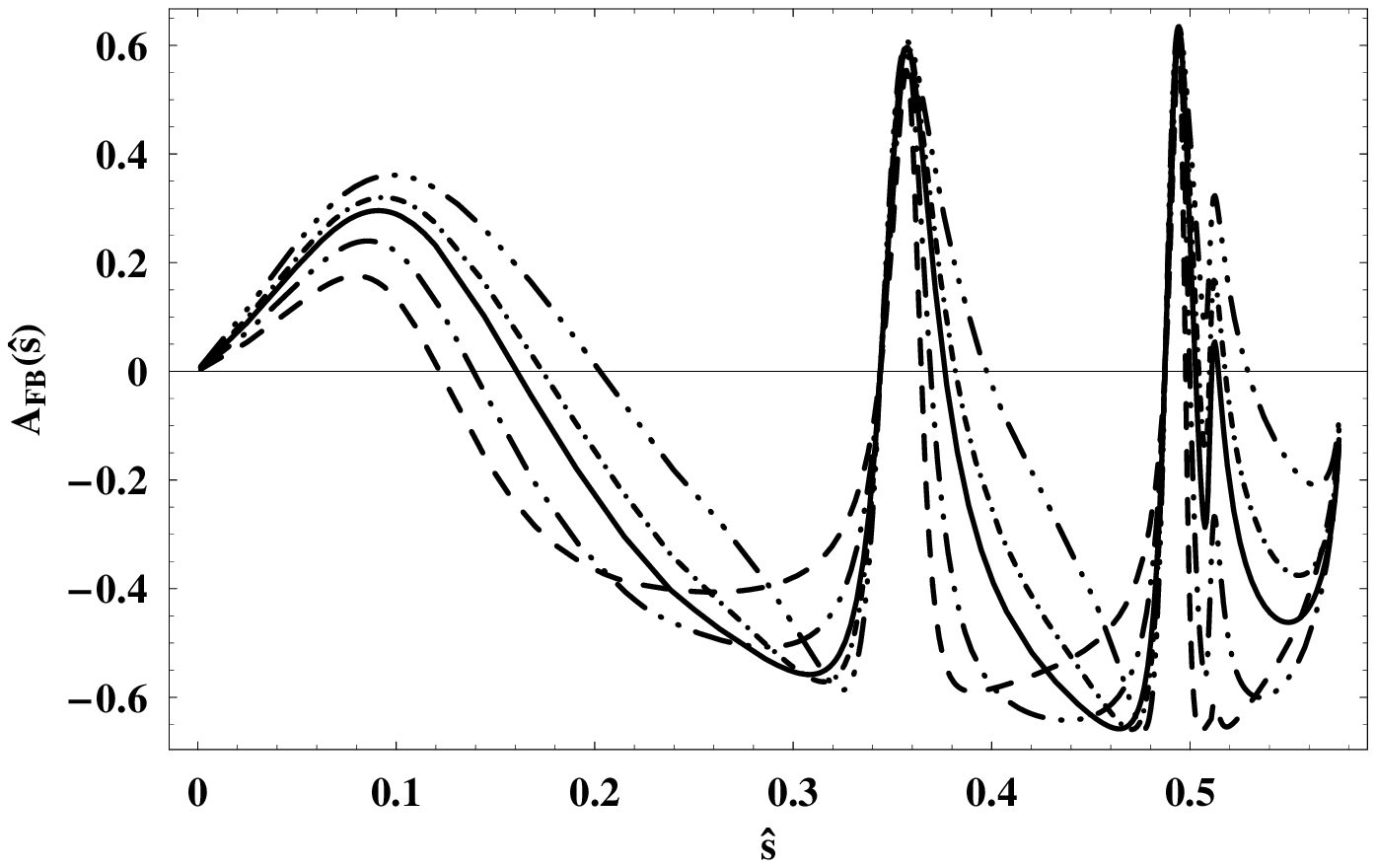} \put (-350,220){(a)} \put
(-100,220){(b)} &  &
\end{tabular}%
\end{center}
\caption{Forward-backward asymmetry for the $B\rightarrow K_{1}\protect\mu %
^{+}\protect\mu ^{-}$ decays as functions of $\hat{s}$ for different values
of $C_{LL(LR)}$. Solid line correspond to SM value,dashed line is for $%
C_{LL(LR)}=-C_{10}$, dashed-dot-dot is for $C_{LL(LR)}=-0.7C_{10}$, dashed
dotted line is for $C_{LL(LR)}=C_{10}$, dashed-triple-dotted is for $%
C_{LL(LR)}=0.7C_{10}$. The coefficients of the other interactions are all
set to zero. }
\label{forward backward asymmetry-CLL}
\end{figure}

\textbf{Switching on} $\mathbf{C}_{RR}$ and $C_{RL}$\textbf{\ along with SM\
operators}

In Fig. 2(a, b) we have shown the dependence of forward-backward asymmetry
on $C_{RR} $ and $C_{RL}$. Fig. 2a give the plot of the $\mathcal{A}_{FB}$
with $\hat{s}$ by using different values of $C_{RR}$ and setting all the
other Wilson Coefficients to zero. By varying the $C_{RR}$ from $-C_{10}$ to
$C_{10} $ in the same way as we did for the $C_{LL}$ in Fig. 1, we have
plotted the $\mathcal{A}_{FB}$ with $\hat{s}$ $\ $in Fig. 2a where, the
legends of the curves are the same as in Fig. 1. One can clearly see that
the zero position of the forward-backward asymmetry is less sensitive to $%
C_{RR}$ compares to the $C_{LL}$ and $C_{LR}$ and the position of the zero
shifts left to the SM\ value from $\hat{s}$ = $0.16$ to $0.12$ when $C_{RR}$
is changed from $-C_{10}$ to $C_{10}$. Again this is contrary to the $B \to
K^{*}\mu^{+}\mu^{-}$ case where is the shift of zero position of $A_{FB}$ is
on the other way.

Similarly Fig. 2b shows the dependency of the zero position of forward
backward asymmetry on different values of $C_{RL}$. It can be seen that when
$C_{RL}$ vary from $-C_{10}$ to $C_{10},$ the zero position of the $\mathcal{%
A}_{FB}$ shifts gradually right to the SM value from $\hat{s}$ = $0.16$ to $%
0.21$ .

\begin{figure}[h]
\begin{center}
\begin{tabular}{ccc}
\vspace{-2cm} \includegraphics[scale=0.6]{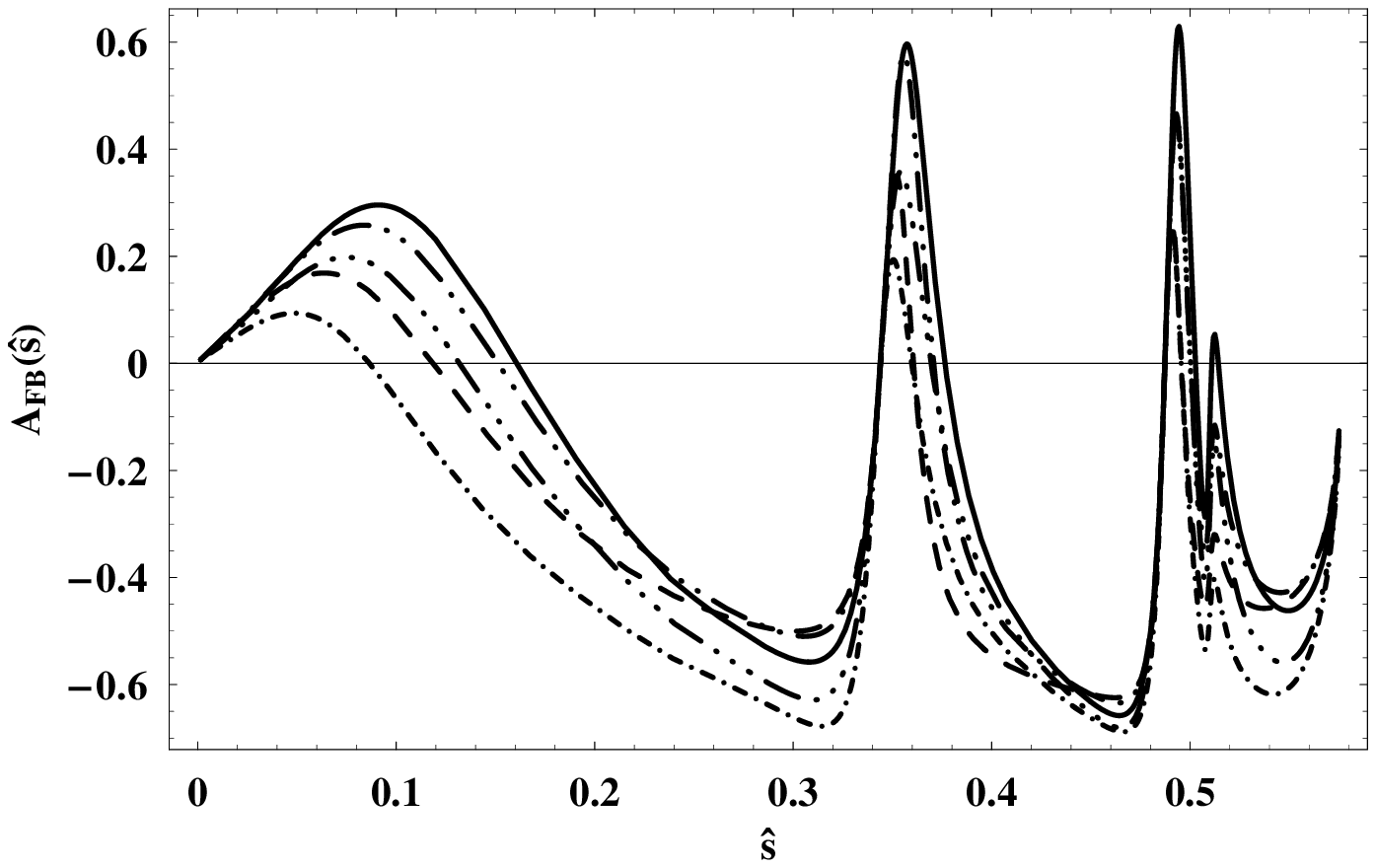}%
\includegraphics[scale=0.6]{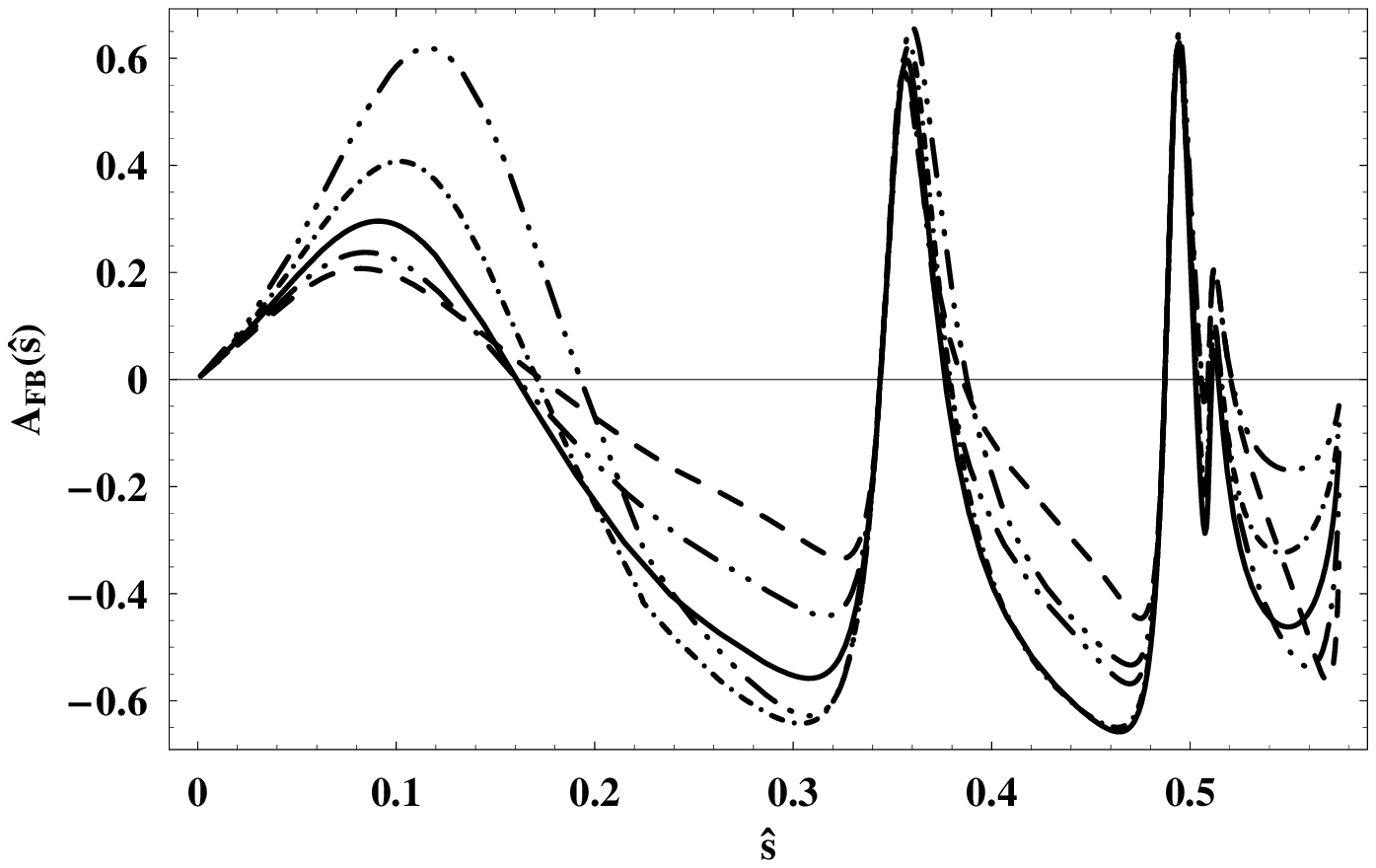} \put (-350,220){(a)} \put
(-100,220){(b)} &  &
\end{tabular}%
\end{center}
\caption{Forward-backward asymmetry for the $B \to K_1 \protect\mu^+\protect%
\mu^-$ decays as functions of $\hat{s}$ for different values of $C_{RR(RL)}$%
. Solid line correspond to SM value,dashed line is for $C_{RR(RL)}=-C_{10}$,
dashed-dot-dot is for $C_{RR(RL)}=-0.7C_{10}$, dashed dotted line is for $%
C_{RR(RL)}=C_{10}$, dashed-triple-dotted is for $C_{RR(RL)}=0.7C_{10}$. The
coefficients of the other interactions are all set to zero. }
\label{forward backward asymmetry-CLL}
\end{figure}

\textbf{Switching only Scalar- Psudoscalar }$\left( C_{LRLR}\text{, }C_{RLLR}%
\text{, }C_{LRRL}\text{, }C_{RLRL}\right) $\textbf{operators along with\ SM
operators}

Fig. (3) shows the behavior of the lepton forward-backward asymmetry for
different NP scalar operators. In the graph we have chosen the value of the scalar and psudoscalar
operators such that they satisfy the constraint
 $R\equiv|C_{LRRL}^{(+)}-C_{RLLR}^{(+)}|^2+|C_{LRRL}^{(-)}-C_{RLLR}^{(-)}|^2\leq0.44$ as provided by
the $\bar{B^{0}_{s}}\to \mu^{+}\mu^{-}$ decay \cite{newref3}. It can be seen from the Eq. (\ref%
{expressions-FBA}) that the contribution from the scalar operators alone is
zero.\ This is quite clear in the graph where the value of $\mathcal{A}_{FB}$
overlap with that of the SM value and this is due to the interference
between the NP scalar operators and that of the SM operators (i.e their
coefficients).

\begin{figure}[t]
\begin{center}
\begin{tabular}{ccc}
\vspace{-2cm} \includegraphics[scale=0.6]{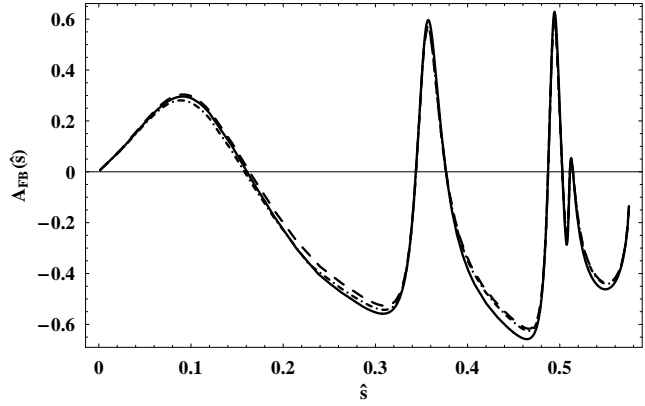} &  &
\end{tabular}%
\end{center}
\caption{Forward-backward asymmetry for the $B\rightarrow K_{1}\protect\mu %
^{+}\protect\mu ^{-}$ decays as functions of $\hat{s}$ for different values
of Scalar and Psudoscalar operators. Solid line correspond to SM
value,dashed line is for $R$=$0.44$ and dashed dotted is for $R$$<0.44$. The
coefficients of the other NP interactions are all set to zero.}
\label{forward backward asymmetry-CLL}
\end{figure}

\textbf{Switching on only Tensor-Axial Tensor}$\left( C_{T}\text{, }%
C_{TE}\right) $\textbf{operators along with\ SM operators}

\begin{figure}[h]
\begin{center}
\begin{tabular}{ccc}
\vspace{-2cm} \includegraphics[scale=0.6]{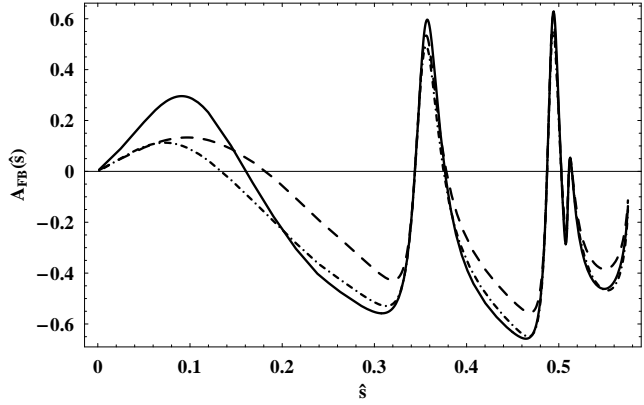} &  &
\end{tabular}%
\end{center}
\caption{Forward-backward asymmetry for the $B\rightarrow K_{1}\protect\mu %
^{+}\protect\mu ^{-}$ decays as functions of $\hat{s}$ for different values
of Scalar and Psudoscalar operators. Solid line correspond to SM
value,dashed line is for $\left|C_{T}\right|^{2}+4\left|C_{TE}\right|^{2}$=$%
1.3$, dashed-dot is for $\left|C_{T}\right|^{2}+4\left|C_{TE}\right|^{2}$=$%
0.9$. The coefficients of the other NP interactions are all set to zero.}
\label{forward backward asymmetry-CLL}
\end{figure}

This is the case where only NP tensor operators are added. It is expected
from Eq. (\ref{expressions-FBA}) that the contribution alone from the tensor
operators to $\mathcal{A}_{FB}$ is zero and Fig. (4) reflects this scenario.
Just like the scalar operators, the non zero value of the forward-backward
asymmetry is due to the interference between the tensor type operator and of
the SM operators and these are $\hat{m_{l}}$ suppressed (c. f. Eq. (\ref%
{expressions-FBA})). The allowed values of new tensor type operators are
restricted to be \cite{newref3}
\begin{equation}
\left|C_{T}\right|^{2}+4\left|C_{TE}\right|^{2}\leqslant 1.3
\end{equation}

In Fig. 4 one can see the $\hat{m_{l}}$ suppression (which is not
negligible) for the value of $A_{FB}(\hat{s})$ in the low $\hat{s}$ region.
Though the value is suppressed but still the shift in the zero position is
quite significant in the low $\hat{s}$ region, which is due to the mixing of
Tensor and SM interactions.\newline
\textbf{Combination of SP, VA and T operators}

\begin{figure}[ht]
\begin{center}
\begin{tabular}{ccc}
\vspace{-2cm} \includegraphics[scale=0.6]{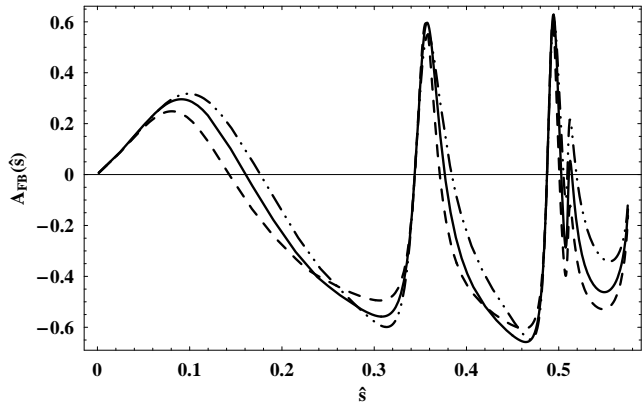} &  &
\end{tabular}%
\end{center}
\caption{Forward-backward asymmetry for the $B\rightarrow K_{1}\protect\mu %
^{+}\protect\mu ^{-}$ decays as functions of $\hat{s}$ for different values
of Vector and Axialvector operators. Solid line
correspond to SM value, dashed line is for VA couplings equal to $-0.3C_{10}$ and dashed-dot-dot lines
are for VA equal to $0.3C_{10}$. Here took the value of SP operators such that they satisfy $R$=$0.44$. The
coefficients of the other NP interactions are all set to zero.}
\label{forward backward asymmetry-CLL}
\end{figure}

Apart from the individual contribution of NP operators and their
interference with the SM operators there is a also a mixing between NP
operators by itself. By looking at the term $X_{SP-VA}$ in Eq. (\ref%
{expressions-FBA}) one can see that it is $\hat{m_{l}}$ suppressed but with
the second term there is a factor of $M_{B}^2$ which will over come this
suppression. This will not only change the zero position of $A_{FB}$ but
also increases or decreases its value compared to SM value depending on the
size and sign of NP couplings. In Fig. 5, we took $R$=$0.44$ and the values of NP vector type
operators is taken to be $0.3C_{10}$ or $-0.3C_{10}$.

Among different mixing terms the most important is the SP and T term. Though
the individual contribution of SP and T to the $A_{FB}$ are not very
significant but their interference term is quite promising. One can see it
from $X_{SP-T}$ term in Eq. ({\ref{expressions-FBA}}) in which there is no
lepton mass suppression. In Fig. 6, we have shown the dependencies of the
zero position of forward-backward asymmetry for different values of SP
couplings. The value of tensor couplings is chosen to be $%
\left|C_{T}\right|^{2}+4\left|C_{TE}\right|^{2}\leqslant 1.3$.

Finally, the contribution from the mixing terms of VA and T is suppressed by
$\hat{m_{l}}$ which can be seen in $X_{VA-T}$ term of Eq. {\ref%
{expressions-FBA}}.

\begin{figure}[ht]
\begin{center}
\begin{tabular}{ccc}
\vspace{-2cm} \includegraphics[scale=0.6]{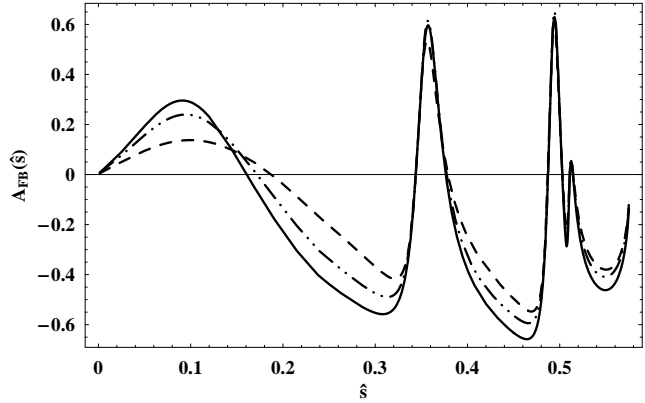} &  &
\end{tabular}%
\end{center}
\caption{Forward-backward asymmetry for the $B\rightarrow K_{1}\protect\mu %
^{+}\protect\mu ^{-}$ decays as functions of $\hat{s}$. Solid line correspond to SM
value,dashed line is for $\left|C_{T}\right|^{2}+4\left|C_{TE}\right|^{2}= 1.3$
and dashed-dot-dot is for $\left|C_{T}\right|^{2}+4\left|C_{TE}\right|^{2}= 0.9$.
Here we kept $R$=$0.44$ and the coefficients of other VA NP interactions are all set to zero.}
\label{forward backward asymmetry-CLL}
\end{figure}

\section{Conclusion:}

The sensitivity of the zero position of the forward backward asymmetry to
the new physics effects is studied here. We showed that the position of the
zero of the forward backward asymmetry shifts significantly from its
Standard Model value both for the size and sign of the vector-vector new
physics operators which are the opposite chirality part of the corresponding
SM operators. The scalar-scalar four fermion interactions have very mild
effects on the zero of the forward-backward asymmetry. The tensor type
interactions shifts the zero position of the forward-backward asymmetry but
these are $\hat{m_{l}}$ suppressed. However, the interference of SP and T
operators gives significant change in the zero position of $\mathcal{A}_{FB}$%
.

In short, our results provide, just as in case of the $B\rightarrow K^{\ast
}l^{+}l^{-}$ process, an opportunity for the straightforward comparison of
the basic theory with the experimental results, which may be expected in
near future for this process.

\section*{Acknowledgements}

The authors would like to thank Profs. Riazuddin and Fayyazuddin for their
valuable guidance and helpful discussions. The authors M. A. P. and M. J. A.
would like to acknowledge the facilities provided by National Centre for
Physics during this work.


\begin{thebibliography}{99}
\bibitem{1} S. L. Glashow, J. Iliopoulos, and L. Maiani, Phys. Rev. D2, 1285
(1970).

\bibitem{2} N. Cabbibo Phys. Rev. Lett. 10, 531 (1963); M. Kobayashi and K.
Maskawa, Prog. Theor. Phys. 49, 652 (1973).

\bibitem{1exp} B. Aubert \textit{et al}. [BABAR Collaboration], Phys. Rev.
\textbf{D70} (2004) 112006 [arXiv: hep-ex/0407003]

\bibitem{2exp} M. Nakao \textit{et al}.[Belle Collaboration],Phys. Rev.
\textbf{D69} (2004) 112001 [arXiv: hep-ex/0402042]

\bibitem{3exp} T. E. Coan \textit{et al}. [CDF Collaboration], Phys. Rev.
Lett. \textbf{84}(2000) 5283 [arXiv: hep-ex/9912057]

\bibitem{4exp} H. Yang \textit{et al}, Phys. Rev. Lett. \textbf{94}(2005)
111802 [arXiv: hep-ex/0412039]

\bibitem{5exp} A. Ishikawa \textit{et al}.[Belle Collaboration],Phys. Rev.
Lett. \textbf{91} (2003) 261601 [arXiv: hep-ex/0308044]

\bibitem{6exp} B. Aubert \textit{et al}. [BABAR Collaboration],Phys. Rev.
\textbf{D73} (2006) 092001 [arXiv: hep-ex/0604007]

\bibitem{7exp} A. Ishikawa \textit{et al}.[Belle Collaboration],Phys. Rev.
Lett. \textbf{96} (2006) 251801 [arXiv: hep-ex/0603018]

\bibitem{8exp} B. Aubert \textit{et al}. [BABAR Collaboration], arXiv:
hep-ex/0804.4412

\bibitem{9exp} B. Aubert \textit{et al}. [BABAR Collaboration], arXiv:
hep-ex/0807.4119.

\bibitem{11} T. M. Aliev, M. K. Cakmak and M. Savci, Nucl. Phys. \textbf{B607%
} (2001) 305 [arXiv:hep-ph/0009133] ; T. M. Aliev, A. Ozpineci, M. Savci and
C. Yuce, Phys. Rev. \textbf{D66}(2002)115006 [arXiv:hep-ph/0208128] ; T. M.
Aliev, A. Ozpineci and M. Savci, Phys. Lett. \textbf{B511} (2001) 49
[arXiv:hep-ph/0103261] ; T. M. Aliev and M. Savci, Phys. Lett. \textbf{B481}%
(2000)275 [arXiv:hep-ph/0003188] ; T. M. Aliev, D. A. Demir and M. Savci,
Phys. Rev. \textbf{D62} (2000) 074016 [arXiv:hep-ph/9912525] ; T. M. Aliev,
C. S. Kim and Y. G. Kim,Phys. Rev. \textbf{D62} (2000) 014026
[arXiv:hep-ph/9910501] ; T. M. Aliev and E. O. Iltan, Phys. Lett. \textbf{%
B451}(1999) 175 [arXiv:hep-ph/9804458] ; C. H. Chen and C. Q. Geng, Phys.
Rev. \textbf{D66} (2002) 034006 [arXiv:hep-ph/0207038] ; C. H. Chen and C.
Q. Geng, Phys. Rev. \textbf{D66}, 014007 (2002) [arXiv:hep-ph/0205306]. G.
Erkol and G. Turan, Nucl. Phys. \textbf{B635} (2002)286
[arXiv:hep-ph/0204219] ; E. O. Iltan, G. Turan and I. Turan, J. Phys.
\textbf{G28}, 307 (2002) [arXiv:hep-ph/0106136] ; T. M. Aliev, V. Bashiry
and M. Savci, JHEP \textbf{0405} (2004) 037 [arXiv:hep-ph/0403282]. W. J.
Li, Y. B. Dai and C. S. Huang, arXiv:hep-ph/0410317 ; Q. S. Yan, C. S.
Huang, W. Liao and S. H. Zhu, Phys. Rev. \textbf{D62} (2000)094023
[arXiv:hep-ph/0004262]. S. R. Choudhury, N. Gaur, A. S. Cornell and G. C.
Joshi, Phys. Rev. \textbf{D68} (2003) 054016 [arXiv:hep-ph/0304084] ; S. R.
Choudhury, A. S. Cornell, N. Gaur and G. C. Joshi, Phys. Rev. \textbf{D69}
(2004) 054018 [arXiv:hep-ph/0307276]. A. Ali, E. Lunghi, C. Greub and G.
Hiller, Phys. Rev. \textbf{D66} (2002) 034002 [arXiv:hep-ph/0112300] ; F.
Kruger and E. Lunghi, Phys. Rev. \textbf{D63} (2001) 014013
[arXiv:hep-ph/0008210]; F.Kruger, L.M. Seghal, N.Sinha, R.Sinha, Phys.Rev.
\textbf{D61 }(2000) 114028; Erratum-ibid. \textbf{D63} (2001) 019901
[arXiv:hep-ph/9907386]; C.S. Kim, Yeong Gyun Kim (Yonsei U), Cai-Dian Lu,
Takuya Morozumi (Hiroshima U), Phys.Rev. \textbf{D62} (2000) 034013 [arXiv:
hep-ph/0001151]; Thorsten Feldmann (RWTH Aachen), Joaquim Matias (UA
Barcelona), JHEP \textbf{0301} (2003) 074 [arXiv: hep-ph/0212158]. Frank Kr%
\"{u}ger, Joaquim Matias, Phys.Rev. \textbf{D71} (2005) 094009 [arXiv:
hep-ph/0502060] ; P. Colangelo, F. De Fazio, R. Ferrandes, T.N. Pham,
Phys.Rev. \textbf{D73} (2006) 115006 [arXiv: hep-ph/0604029]; E. Lunghi, J.
Matias, JHEP 0704:058,2007 [arXiv: hep-ph/0612166] ; U. Egede, T. Hurth, J.
Matias, M. Ramon, W. Reece, JHEP 0811:032,2008 [arXiv:hep-ph/0807.2589];
Wolfgang Altmannshofer, Patricia Ball, Aoife Bharucha, Andrzej J. Buras,
David M. Straub, Michael Wick, JHEP 0901:019,2009 [arXiv:hep-ph/0811.1214];
Ulrik Egede, Tobias Hurth, Joaquim Matias, Marc Ramon, Will Reece, [arXiv:
hep-ph/1005.0571].

\bibitem{new1} G. Burdman, Phys. Rev. \textbf{D57} (1998) 4254

\bibitem{new2} A. Ali, P. Ball, L. T. Handoko and G. Hiller, Phys. Rev.
\textbf{D61}(2000) 074024 [arXiv:hep-ph/9910221].

\bibitem{nw1} M. Beneke, Th. Feldmann, D. Seidel (Aachen), Nucl.Phys.\textbf{%
B612}:25-58,2001 [arXiv:hep-ph/0106067]. A. Khodjamirian, Th. Mannel, A.A.
Pivovarov, Y.-M. Wang [arXiv:hep-ph/1006.4945].

\bibitem{new3} T. Goto et al., Phys. Rev. D \textbf{55} (1997) 4273; \ T.
Goto, Y.\ Okada and Y. Shimizu, Phys. Rev.\ D \textbf{58} (1998) 094006; S.
Bertolini, F. Borzynatu, A. Masiero and G.\ Ridolfi, Nucl.\ Phys. B \textbf{%
353} (1991) 591.

\bibitem{new4} S. Fukae, C. S. Kim, T. Morozumi, and T. Yoshikawa, Phys.
Rev. \textbf{D62} (1999) 074013.

\bibitem{new5} T. M. Aliev, C. S. Kim and Y. G. Kim, Phys. Rev. \textbf{D62}
(2000) 014026.

\bibitem{new6} S. Fukae, C. S. Kim and T. Yoshikawa, Phys. Rev. \textbf{D61}
(2000) 074015, T. M. Aliev, A. Ozpineci and M. Savci, Phys. Lett. \textbf{%
B511} (2001) 49.

\bibitem{newref1} A. Ishikawa \textit{et al}., Phys. Rev.\ Lett. \textbf{96}
(2006) 251801; J. T. Wei \textit{et al}. [BELLE Collaboration], Phys. Rev.
Lett. \textbf{103} (2009) 171801.

\bibitem{newref2} B. Aubert \textit{et al}. [BABAR Collaboration], Phys.
Rev. \textbf{D 73} (2006) 092001, Phys.\ Rev. \textbf{D 79} (2009) 031102.

\bibitem{newref3} A. \ K. Alok \textit{et al}., JHEP \textbf{1002},
(2010)053.

\bibitem{newr11} H. Hatanaka, K.-C. Yang, Phys. Rev. \textbf{D77} (2008)
094023.

\bibitem{newr12} K.-C. Yang, Phys. Rev. \textbf{D78} (2008) 034018; S. R.
Choudhary, A. S. Cornell, N. Gaur, Eur. Phys. J. \textbf{C58} (2008) 251; V.
Bashiry, JHEP 0906 (2009) 062; V. Bashiry, K.Azizi, arXiv: hep-ph/0903.1505;
M.\ Ali Paracha, Ishtiaq Ahmed and M. Jamil\ Aslam, Eur. Phys. J.\textbf{C52}
(2007) 967; Ishtiaq Ahmed, M. Ali Paracha and M.Jamil Aslam, Eur. Phys. J.
\textbf{C54}(2008) 591; A. Saddique, M. J. Aslam and C. D. Lu, Eur.Phys.J.
\textbf{C56} (2008) 267; J. P. Lee, Phys. Rev. \textbf{D74} (2006) 074001;
M. J. Aslam and Riazuddin, Phys.Rev. \textbf{D72} (2005) 094019; M. J.
Aslam, Eur.Phys.J. \textbf{C49} (2007) 651; K.-C. Yang, Nucl. Phys. \textbf{%
B776} (2007) 187; H. Dag, A. Ozpineci, M. T. Zeyrek, arXiv:1001.0939.

\bibitem{new8} J.Dickens, V.Gibon, C.Lazzeroni and M.Patel,
CERN-LHCB-2007-038, J.Dickens, V.Gibon, C.Lazzeroni and M.Patel,
CERN-LHCB-2007-039.

\bibitem{new9} B.Aubert $et al$ [BABAR Collaboration], Phys.Rev.Lett 91
(2003) 221802 [hep-ex/0308042] ; A.Ishikawha $et al$, Phys.Rev.Lett.
96(2006) 092001 [hep-ex/0604007]; B.Aubert $etal$, arXiv:0804.4412 ;
I.Adachi $etal$, arXiv: 0810.0335.

\bibitem{new10} M-O Bettler \textit{et al.}, for LHCb Collaboration,
CERN-LHCB-CONF-2009-038, LPHE-2009-05, arXiv: 0910.0942.

\bibitem{23} M.\ Ali Paracha, Ishtiaq Ahmed and M. Jamil\ Aslam, Eur. Phys.
J.\textbf{C52}, 967-973 (2007)

\bibitem{PDG} C.~Amsler \textit{et al.} [Particle Data Group],
Phys.\ Lett.\ B \textbf{667}, 1 (2008).

\bibitem{b to s in theory 3} C.S. Kim, T. Morozumi, A.I. Sanda, Phys. Lett.
B \textbf{218} (1989) 343.

\bibitem{b to s in theory 4} X.~G.~He, T.~D.~Nguyen and R.~R.~Volkas,
Phys.\ Rev.\ D \textbf{38} (1988) 814.

\bibitem{b to s in theory 5} B. Grinstein, M.J. Savage, M.B. Wise, Nucl.
Phys. B \textbf{319} (1989) 271.

\bibitem{b to s in theory 6} N.~G.~Deshpande, J.~Trampetic and K.~Panose,
Phys.\ Rev.\ D \textbf{39} (1989) 1461.

\bibitem{b to s in theory 7} P.~J.~O'Donnell and H.~K.~K.~Tung,
Phys.\ Rev.\ D \textbf{43} (1991) 2067.

\bibitem{b to s in theory 8} N.~Paver and Riazuddin,
Phys.\ Rev.\ D \textbf{45} (1992) 978.

\bibitem{b to s in theory 9} A. Ali, T. Mannel and T. Morozumi, Phys. Lett. B%
\textbf{273} (1991) 505.

\bibitem{b to s 1} D.~Melikhov, N.~Nikitin and S.~Simula,
Phys.\ Lett.\ B \textbf{430} (1998) 332 [arXiv:hep-ph/9803343].

\bibitem{b to s 2} J.~M.~Soares, Nucl.\ Phys.\ B \textbf{367} (1991) 575.

\bibitem{b to s 3} G.~M.~Asatrian and A.~Ioannisian,
Phys.\ Rev.\ D \textbf{54} (1996) 5642 [arXiv:hep-ph/9603318].

\bibitem{NF charm loop} J.~M.~Soares,
Phys.\ Rev.\ D \textbf{53} (1996) 241 [arXiv:hep-ph/9503285].

\bibitem{17} Konstantin Chetyrkin, Mikolaj Misiak, Manfred Munz, Phys.Lett. B%
\textbf{400 }(1997) 206-219; Erratum-ibid. B425 (1998) 414
[arXiv:hep-ph/9612313] ; Christoph Bobeth, Mikolaj Misiak, Joerg Urban,
Nucl.Phys. B\textbf{574} (2000) 291-330 [arXiv:hep-ph/9910220] ; A.
Ghinculov, T. Hurth, G. Isidori, Y.-P. Yao, Nucl.Phys. B\textbf{685} (2004)
351-392 [arXiv:hep-ph/0312128].

\bibitem{16} Altug Arda and Muge Boz, Phys. Rev. D \textbf{66}, 075012
(2002).

\bibitem{naveen05} A. S. Cornell, Naveen Gaur, Sushil K. Singh,
hep-ph/0505136.
\end{thebibliography}
\end{document}